\documentclass[%
 reprint,
 amsmath,amssymb,
 aps,
pra
]{revtex4-2}

\usepackage{graphicx}% Include figure files
\usepackage{dcolumn}% Align table columns on decimal point
\usepackage{bm}% bold math
\usepackage{tabularx}
\usepackage[hidelinks]{hyperref}% add hypertext capabilities
\begin{document}

\preprint{APS/123-QED}

\title{Error-detected coherence metrology of a dual-rail encoded fixed-frequency multimode superconducting qubit}

\author{James Wills}%
\email{jwills@oqc.tech}
\author{Mohammad Tasnimul Haque}
\author{Brian Vlastakis}%
\email{bvlastakis@oqc.tech}

\affiliation{Oxford Quantum Circuits, Thames Valley Science Park, Shinfield, Reading, United Kingdom, RG2 9LH}

\date{\today}

\begin{abstract}
Amplitude damping is a dominant source of error in high performance quantum processors. A promising approach in quantum error correction is erasure error conversion, where errors are converted into detectable leakage states. Dual-rail encoding has been shown as a candidate for the conversion of amplitude-damping errors; with unique sensitivities to noise and decoherence sources. Here we present a dual-rail encoding within a single fixed-frequency superconducting multimode transmon qubit. The three island, two junction device comprises two transmonlike modes with a detuning of 0.75-1 GHz, in a coaxial circuit QED architecture. We show the logical bit-flip and phase-flip error rates are more than one order of magnitude lower than the physical error rates, and demonstrate stability and repeatability of the architecture through an extended measurement of three such devices. Finally, we discuss how the error-detected subspace can be used for investigations into the fundamentals of noise and decoherence in fixed-frequency transmon qubits.

\end{abstract}

\maketitle

\section{\label{sec:introduction}Introduction}
Amplitude damping as a result of energy relaxation is a significant source of error in superconducting qubits for use in quantum computing, characterized by the decay constant conventionally known as $T_1$. Significant research efforts exist to try to increase qubit $T_1$ via nanofabrication methods \cite{siddiqi_engineering_2021, murray_material_2021}, novel qubit designs such as fluxonium \cite{manucharyan_fluxonium_2009, nguyen_high-coherence_2019, somoroff_millisecond_2021}, and through engineering of the microwave environment and Purcell filtering \cite{reed_fast_2010, sete_quantum_2015, bronn_reducing_2015, bronn_broadband_2015, sunada_fast_2022}.

Quantum error correction exists as a method of reducing errors in physical qubits via the redundant encoding of information across a large array of noisy physical qubits \cite{steane_simple_1996, shor_fault-tolerant_1997, fowler_surface_2012, bravyi_high-threshold_2024}. Whilst early demonstrations of surface codes have shown modest improvement in logical errors over physical error rates \cite{acharya_suppressing_2023, acharya_quantum_2025}, the hardware overheads required to sufficiently reduced errors to rates essential for useful quantum computation are dauntingly large.

Novel encoding and error-detection mechanisms have been suggested to increase efficiency in error-correction protocols and reduce these hardware requirements. Notably, erasure error-conversion \cite{wu_erasure_2022, kubica_erasure_2023}, defined by converting noise into heralded leakage states, has been theoretically shown to have more favorable error thresholds \cite{gu_fault-tolerant_2023, gu_optimizing_2024, sahay_high-threshold_2023, chang_surface_2024}. As such, it may be possible to reach lower logical qubit error rates for the same number of noisy physical qubits.

One method of implementing an erasure error-conversion is dual-rail encoding, where quantum information is encoded across the subspace of two physical entities. This encoding has been demonstrated in superconducting quantum devices in both superconducting cavities \cite{teoh_dual-rail_2023, chou_demonstrating_2023, koottandavida_erasure_2023, graaf_mid-circuit_2024}, as well as flux tunable transmon qubits \cite{levine_demonstrating_2024}. In each case, the encoding and error-detection protocols show an improvement in the error-rates of the logically encoded qubit compared to the physical constituents, demonstrating the viability of this architecture.

In this paper, we present an implementation of a dual-rail encoding in a fixed-frequency multimode transmon, within a coaxial circuit QED architecture. We demonstrate all-microwave control and readout of the logical qubit, utilizing an end-of-line error-detection protocol, and characterize idling error-rates of the encoded subspace. The error-detected logical coherence metrics are shown to have an order of magnitude improvement in bit-flip and phase-flip error rates as compared to the constituent physical qubit modes. In addition, we show how the logical subspace can be used as a tool to probe the fundamentals of the noise and decoherence mechanisms in superconducting circuits. It is vital to understand these sources of error more deeply in order to achieve the goal of a fault tolerant quantum computer.

\section{\label{sec:device}Dual-Rail Encoding in a Fixed Frequency Multi-mode Transmon Qubit}

\begin{figure}
\includegraphics[width=0.48\textwidth]{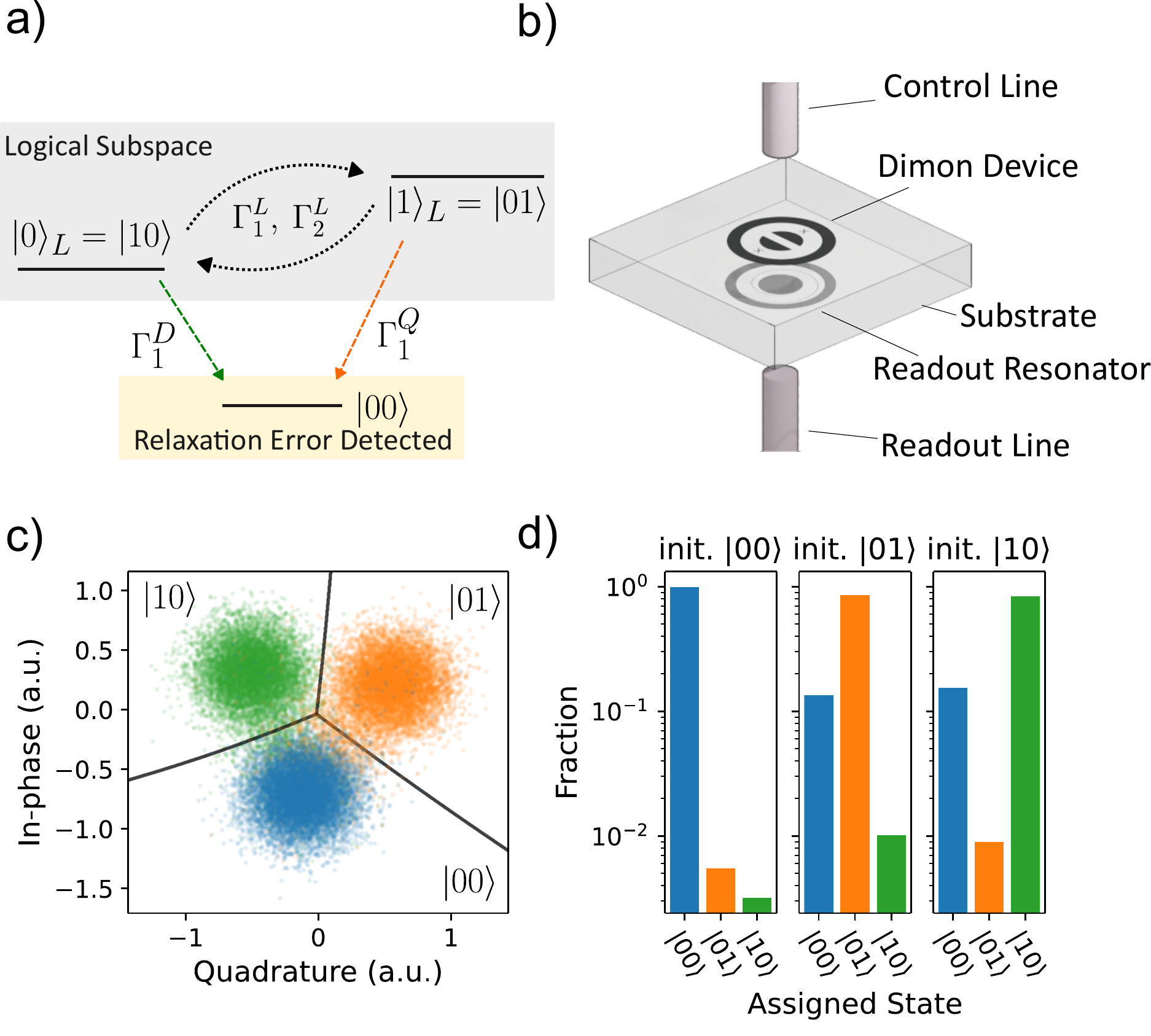}
\caption{\label{fig:device}Device description. (a) Energy level diagram of the dimon device. Logical subspace highlighted in gray, and excitation and relaxation error-detected states highlighted in yellow. Orange (green) arrows indicate energy level transitions, and labeled transition rates, following relaxation events of the dimon quadrupole (dipole) mode. (b) Schematic of the multimode qubit in a coaxial circuit QED architecture unit-cell. (c) Integrated readout signal of the end-of-line (EOL) logical state discrimination readout protocol. Distributions correspond to the multimode qubit prepared in the $|00 \rangle$ (blue), $|01 \rangle$ (orange) and $|10 \rangle$ (green) states. Black lines indicated state discrimination boundaries as determined by the fitted Gaussian mixture model (GMM). (d) Physical state assignment fractions with DDQ (Q1) initialized in the $|00 \rangle$ (left), $|01 \rangle$ (center), and $|10 \rangle$ (right) states.}
\end{figure}

Where a conventional transmon qubit consists of two superconducting islands and one Josephson junction \cite{koch_charge_2007}, the multimode transmon consists of three superconducting islands, and two Josephson junctions \cite{wills_spatial_2022}. This additional island and junction adds a degree of freedom to the circuit, and subsequently a second transmonlike mode. These devices have been widely explored for a number of use cases such as photon shot noise suppression and tunable coupling \cite{srinivasan_tunable_2011, hoffman_coherent_2011, zhang_suppression_2017}, mode-selective coupling and entanglement generation \cite{finck_suppressed_2021, wills_multi-mode_2022, heya_randomized_2025, li_realization_2024}, and Purcell protection and fast readout \cite{dassonneville_fast_2020}.

We construct this device in a coaxial architecture \cite{rahamim_double-sided_2017, patterson_calibration_2019, spring_high_2022}, comprising of the multimode device, and lumped element resonator fabricated on opposite sides of a low loss dielectric substrate, as shown in Fig. \ref{fig:device} (a). Control and readout signals are delivered by capacitively coupled coaxial lines above and below the device. In this implementation, there is no galvanic coupling to the device or substrate, nor is there any requirement for flux tuning. In addition, the device, control, and readout circuitry occupies the same physical footprint of the conventional coaxial qubit \cite{rahamim_double-sided_2017, wills_spatial_2022}, demonstrating the hardware-efficiency of this extensible architecture.

The device consists of two transmonlike modes we label as D and Q, due to the dipolelike and quadrupolelike polarisations of the electric fields that the modes best couple to respectively. We describe these modes as the \textit{physical} modes of the system.

The Hamiltonian describing the circuit in the harmonic oscillator basis is given by,
\begin{equation}
\begin{split}
    \frac{\hat{H}}{\hbar} & = \sum_{i = D, Q} \omega_i \hat{a}^\dagger_i \hat{a}_i - \frac{\alpha_i}{2} \times \hat{a}^\dagger_i \hat{a}_i (\hat{a}^\dagger_i \hat{a}_i - 1) \\
    & -\eta \prod_{i = D, Q}\hat{a}^\dagger_i \hat{a}_i,
\end{split}
\end{equation}
where $\omega_i$ is the transition frequency, $\alpha_i$ is the anharmonicity, and $a^{(\dagger)}_i$ is the annihilation (creation) operator for each mode $i$. The coupling between each mode is purely longitudinal with strength described by the cross-Kerr shift $\eta$. Since this device has two modes, it is known as a dimon \cite{hazra_engineering_2020, hazra_benchmarking_2025}. Each physical mode of the dimon behaves in the same way as a standard transmon, with transition frequencies and anharmonicities summarized in Table \ref{table:1}. Each mode can be addressed via the same coaxial control line shown in Fig. \ref{fig:device} (a), and manipulated via a microwave signal.

The energy level diagram of the dimon is shown in Fig. \ref{fig:device} (a). The states are labeled by $|mn \rangle$, corresponding to $m (n)$ excitations in the D (Q) mode. We implement a dual-rail encoding by encoding a logical qubit in the single-excitation subspace of the energy level structure, such that a logical 0 ($|0 \rangle _L$) corresponds to the $|10 \rangle$ state, and a logical 1 ($|1 \rangle _L$) corresponds to the $|01 \rangle$ state. Notably, there is no direct single photon decay channel from the $|01 \rangle$ state to the $|10 \rangle$ state, due to the different symmetries of the modes. We name this method of encoding, a dual-rail encoded dimon qubit (DDQ).

Amplitude damping, or energy relaxation, will manifest as either the $|10 \rangle$ or $|01 \rangle$ state decaying to the $|00 \rangle$ state, at a rate given by $\Gamma_{1}^{D(Q)} = 1 / T_{1}^{D(Q)}$, where $T_1^{D(Q)}$ is the energy relaxation time of the D (Q) mode. The dual-rail encoding converts this error to leakage outside of the logical subspace, described as an erasure error. Any detectable population of the $|00\rangle$ state will flag that an error has occurred.

The DDQ offers a fixed frequency approach to erasure error detection within the same hardware-efficient physical footprint of the coaxial circuit QED architecture unit-cell. In addition, the need for flux tuning or galvanic coupling to the substrate that can introduce an unwanted noise source or crosstalk and additional calibration parameters is negated.

As shown in Fig. \ref{fig:device} (b), there is a lumped element LC resonator on the opposing side of the substrate. Due to the symmetry of the device, this resonator is transversely coupled to the Q mode of the dimon by a Hamiltonian term of the form $\hat{H}_{QR}/\hbar = g_{QR}(\hat{a}^\dagger_Q + \hat{a}_Q)(\hat{a}^\dagger_R + \hat{a}_R)$, where $\hat{a}^{(\dagger)}_R$ is the annihilation (creation) operator of the readout resonator, and $g_{QR}$ is the transverse coupling strength. Whilst there is no direct coupling between the D mode and the resonator, it does inherit a state-dependent frequency shift due to the longitudinal coupling to the Q mode. In the dispersive limit ($g_{QR} \ll \omega_R - \omega_Q$), the state-dependent frequency shifts $\chi_{Q(D)R}$ between the Q (D) mode and the readout resonator are given by,
\begin{equation}
\begin{split}
    \chi_{QR}  \approx \alpha_Q &\left(\frac{g_{QR}}{\omega_R - \omega_Q}\right)^2, \\
    \chi_{DR}  \approx \eta & \left(\frac{g_{QR}}{\omega_R - \omega_Q}\right)^2, 
\end{split}
\end{equation}
where $\omega_R$ is the frequency of the resonator, and $2\chi_{D(Q)R}$ describes the full frequency shift of the resonator.

We perform an end-of-line (EOL) logical state distinction, designed to maximally distinguish the three states of the dimon. To achieve this, we drive the readout resonator at a frequency of $\omega_R - (\chi_{QR} + \chi_{DR})/2$. We plot the measured signal in the IQ-plane as a function of the initial state of the dimon in Fig. \ref{fig:device} (c), which shows three separate circular distributions. A Gaussian mixture model (GMM) classifier is fit to a training set of 10k state preparation shots in order to build a model to classify subsequent measurement data and assign physical states, the boundaries of which are denoted by the solid lines in Fig. \ref{fig:device} (c). We show an example of the assignment probabilities in Fig. \ref{fig:device} (d).

With this EOL readout method, we are able to perform a number of  experiments using the logical encoding of the DDQ, and detect the shots in which there has been leakage outside of the computational subspace ($P(|00\rangle)$. We can postselect the result to remove this fraction of results, and renormalize the remaining $|01\rangle$ and $|10\rangle$ state populations to obtain the error-detected logical state probabilities, 
\begin{equation}
\begin{split}
    P(|0\rangle_L) = \frac{P(|10\rangle)}{P(|01) \rangle + P(|10\rangle)}, \\
    P(|1\rangle_L) = \frac{P(|01\rangle)}{P(|01) \rangle + P(|10\rangle)}. 
\end{split}
\end{equation}

% \begin{equation}
% \begin{split}
%     P(|0\rangle_L) = P(|10\rangle \:\vert\: \overline{|00\rangle}), \\
%     P(|1\rangle_L) = P(|01\rangle \:\vert\:  \overline{|00\rangle}). 
% \end{split}
% \end{equation}

In this work, we measure a set of 3 fixed-frequency dimon qubits, with frequency and physical mode coherence parameters described in Table \ref{table:1} in Appendix \ref{sec:device_params}. The physical mode frequencies range from $\omega_D/2\pi = 4.24 - 4.71$ GHz ($\omega_Q/2\pi = 5.28 - 5.47$ GHz), with a range of mode detunings $\Delta/2\pi = (\omega_Q - \omega_D)/2\pi = 0.76 - 1.05$ GHz, larger than previous demonstrations of dual-rail encoding in superconducting qubits. The device is fabricated through methods described in \cite{acharya_integration_2024}.

\section{\label{sec:coherence_stats}Error-Detected Coherence Metrics}

\begin{figure*}[t]
\includegraphics[width=1\textwidth]{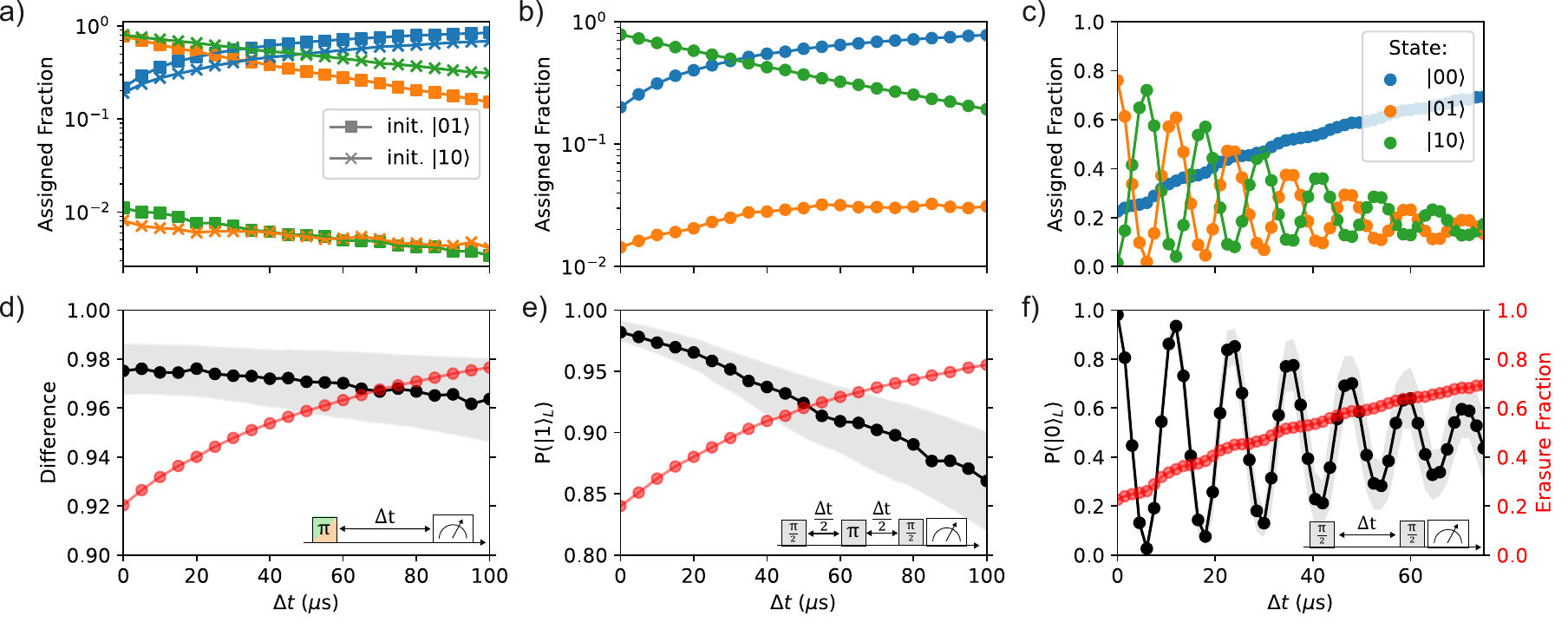}
\caption{\label{fig:coherence_traces}Error-detected coherence metric traces, extracted from 200 repeated measurements of Q1. (a-c) Fraction of states assigned to the $|00 \rangle$ (blue), $|01 \rangle$ (orange), $|10 \rangle$ (green) in measurements of the (a) logical bit-flip rate, (b) logical Hahn-echo and (c) logical Ramsey decay metrics. (d-f) Error-detected logical state populations (black) postselected to remove amplitude damping events, resulting in (d) logical bit-flip rate, (e) logical phase-flip rate and (f) logical Ramsey decay rate. Y-axis in (d) showing difference between $P(|1\rangle _L | \mbox{init.} |1\rangle_L)$ and $P(|1\rangle _L | \mbox{init.} |0\rangle_L)$, as outlined in main text. Shaded area indicates $1\sigma$ deviation. Fraction of physical measurements classified as having an erasure error event shown in red. (Inset) Pulse sequence of the measurement for (d) bit-flip, (e) phase-flip and (f) Ramsey coherence measurements. Physical mode gates in green/orange, and logical qubit gates shown in gray.}
\end{figure*}

\begin{figure}[t]
\includegraphics[width=0.45\textwidth]{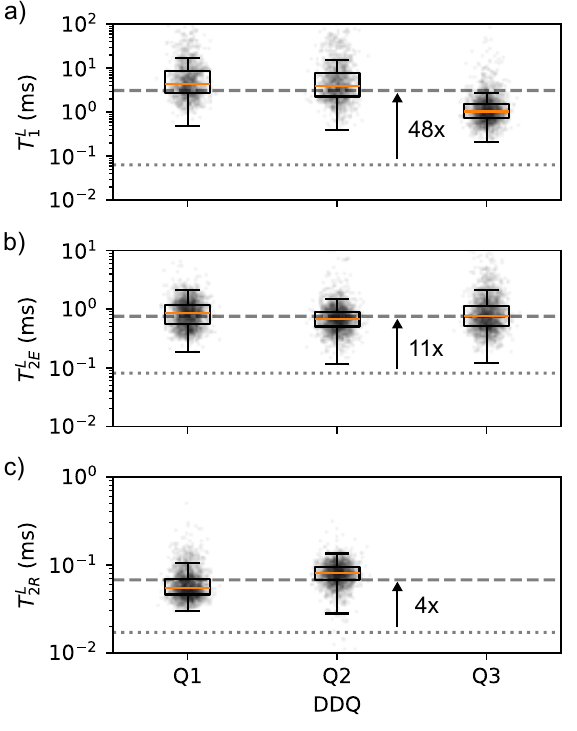}
\caption{\label{fig:coh_stats}Logical coherence metric distributions. Boxplots of the fitted values of (a) logical $T_1$ ($T_1^L$), (b) logical Hahn-echo ($T_{2E}^L$), and (c) logical Ramsey ($T_{2R}^L$), of Q1, Q2 and Q3. Individual trace fits shown as semi-transparent black circles. Horizontal black dashed line indicates mean of the median values of Q1, Q2 and Q3 for each metric. Horizontal dotted line indicates mean of the medians of corresponding non-error-detected physical mode coherence metrics ((a) $T_1$, (b) Hahn-echo $T_{2E}$, (c) Ramsey ($T_{2R}$). Arrows labeled with ratio of logical coherence metric to physical coherence metric.}
\end{figure}

Using our logical state manipulation and readout protocols, we now characterize the logical error-rates of the DDQ: the bit-flip error rate, phase-flip error rate, and Ramsey coherence. 

The logical bit-flip rate is characterized by the rate at which the $|1 \rangle_L$ population transfers to the $|0\rangle_L$ state, and vice-versa. We measure this by preparing the DDQ in each of the two logical states, waiting for a delay time $\Delta t$, and performing an EOL logical state distinction readout, as shown in the inset of Fig. \ref{fig:coherence_traces} (d). Using the readout classifier previously described in Section \ref{sec:device}, each measurement shot is classified into one of the three DDQ states. The logical bit-flip probability is calculated as,  
\begin{equation}\label{eq:bit_flip_probability}
    P(\mbox{bit-flip}) = \frac{1}{2} (P(|1\rangle_L |  \mbox{init.} |0\rangle_L) + P(|0\rangle_L |  \mbox{init.} |1\rangle_L))
\end{equation}

Following the methodology of \cite{levine_demonstrating_2024}, we measure $P|1 \rangle_L$ after initializing the DDQ in the $|01 \rangle$ and $|10 \rangle$ states, and plot the difference between the two error-detected state populations as a function of measurement delay $\Delta t$, shown in Fig. \ref{fig:coherence_traces} (d).

Where a conventional measurement of a superconducting transmon qubit energy relaxation rate (or $T_1$ decay) typically shows an exponential decay, shown in Fig. \ref{fig:coherence_traces} (a), the error-detected logical bit-flip measurement shows a more complex second-order type decay profile. At long time scales, in the absence of mid-circuit erasure error-checks, the logical bit-flip error is dominated by second order relaxation and subsequent re-excitation events. For short timescales relevant for algorithmic implementations and error-correction cycles, the decay profile can be approximated to be linear. To obtain the short circuit depth bit-flip rate, we fit the decay up to $30$ $\mu s$ to a linear slope with constant offset to extract the logical bit-flip rate $\Gamma_1^L$, and define the error-detected logical qubit $T_{1}^L = 1/\Gamma_{1}^L$.

To measure the logical phase-flip rate, we implement a Hahn-echo sequence. We prepare the dimon in the superposition state $|\psi \rangle = (|01\rangle + |10\rangle)/\sqrt{2}$, wait a delay time $\Delta t/2$, apply a refocusing pulse, and project the state back to the logical $Z$ axis after a second delay period of $\Delta t/2$, as shown in the pulse sequence in the inset of Fig. \ref{fig:coherence_traces} (e). Once again, we perform a logical state distinction EOL measurement, and show the DDQ states as a function of delay time, $\Delta t$, in Fig. \ref{fig:coherence_traces} (b). We show the postselected state probability $P(|1\rangle_L)$, along with the erasure fraction in Fig. \ref{fig:coherence_traces} (e).

As with the bit-flip measurement, we observe a non-exponential decay in the error-detected measurement. The phase-flip error-rate for a circuit with short depth can be obtained by fitting the decay up to 30 $\mu s$ with a linear decay. From this, we extract a logical phase-flip rate $\Gamma_{2}^L = 1/T_{2E}^L$. We discuss the effect of noise sources on coherence further in Section \ref{sec:noise}.

Finally, we measure the Ramsey coherence by preparing the DDQ in a superposition state $|\psi \rangle = (|01\rangle + |10\rangle)/\sqrt{2}$, allowing the state to evolve for time $\Delta t$, and projecting back onto the measurement axis, as shown in the pulse sequence in Fig. \ref{fig:coherence_traces} (f). We add a virtual detuning of $\Delta f \approx 75$ kHz by modifying the phase, $\phi$, of the final projection gate such that $\phi = 2 \pi \Delta f \Delta t$, inducing an oscillation between the $|01\rangle$ and $|10\rangle$ states. As previously, we show the measured physical state populations in Fig. \ref{fig:coherence_traces} (c), and postselected logical state probability $P(|0\rangle_L)$, as well as erasure fraction, in Fig. \ref{fig:coherence_traces} (f).

Unlike the bit-flip and phase-flip measurements, the Ramsey decay follows an exponentially decaying oscillation of the form, 
\begin{equation}\label{eq:ramsey_decay}
    P(|0\rangle_L) = A\exp{(-\Delta t/T_{2R}^L)}\cos(2 \pi \Delta f \Delta t + \phi_0) + C,
\end{equation}
where $T_{2R}^L$ is the extracted Ramsey decay constant, $\Delta f$ is frequency detuning, and $A$, $C$ and $\phi_0$ are amplitude and phase offsets.

To build statistics and gain insights into the stability of the error-detected logical qubit coherence metrics, we interleave and repeat measurements of the bit-flip, phase-flip, and Ramsey coherence as previously described on a three DDQ device over the course of over 50 hours (1750 total repetitions each). Q2 and Q3 are measured simultaneously, whilst Q1 is measured independently over a different 50 hour period. We show boxplots of the extracted coherence parameters, $T_1^L$, $T_{2E}^L$, and $T_{2R}^L$ in Fig. \ref{fig:coh_stats} (Q3 Ramsey coherence was unable to be measured due to instability in one of the physical modes).

From these extended measurements, we extract median values of $T_1^L = 3.12$ ms, $T_{2E}^L = 0.76$ ms, and $T_{2R}^L = 0.07$ ms, with individual DDQ device metrics shown in Table \ref{table:1}. In Fig. \ref{fig:coh_stats}, we also show the median physical coherence parameters. The measured error-detected coherence metrics correspond to 48x improvements in bit-flip rates, 11x improvement in phase-flip rate, and 4x improvement in Ramsey coherence times. In addition to the qubit coherence metrics being greatly improved over a physical qubit, these metrics remain stable over the course of the measurement period, as shown in Fig. \ref{fig:coh_time_traces} in Appendix \ref{sec:time_series_data}. Validating the reproducibility of the improved logical coherence metrics across multiple devices with a range of frequencies is a significant step to demonstrate the viability of the DDQ platform for future hardware integrations.

In addition to the error-detected coherence metrics, we measure the erasure rate, obtained from the $|00 \rangle$ state population. This is indicated by the red trace, shown in Fig. \ref{fig:coherence_traces} (d-f). We fit this to the form,
\begin{equation}\label{eq:erasure_decay}
    P(|00\rangle) = \left(1 - \exp\left(-\frac{\Delta t}{T_{erasure}}\right)\right) + D,
\end{equation}
where $1/T_{erasure} = \Gamma_{erasure}$ is the rate of leakage out of the computational subspace, and $D$ is an offset due to erasure events occurring during the EOL readout. We show the extracted erasure rate, as well as physical mode relaxation rates in the time series plot in Fig. \ref{fig:coh_time_traces}, and observe it qualitatively follows the same fluctuations, as expected.

The erasure rate is a important metric for dual-rail encoded qubits, as it informs both the expected leakage rate during error-correction cycles, as well as sampling overhead in postselected quantum algorithm applications.

\section{\label{sec:noise}Noise and Decoherence}

In this implementation of dual-rail encoding, the coherence of the logical qubit is affected by fluctuations in the detuning between the physical modes of the dimon ($\delta \Delta = \delta \omega_Q - \delta \omega_D$), in contrast to a conventional transmon qubit implementation where the coherence is sensitive to fluctuations in the transition frequency $\delta \omega$ \cite{koch_charge_2007}. As such, the DDQ possesses unique sensitivities and insensitivities to physical noise sources.

\begin{figure*}[t]
\includegraphics[width=0.98\textwidth]{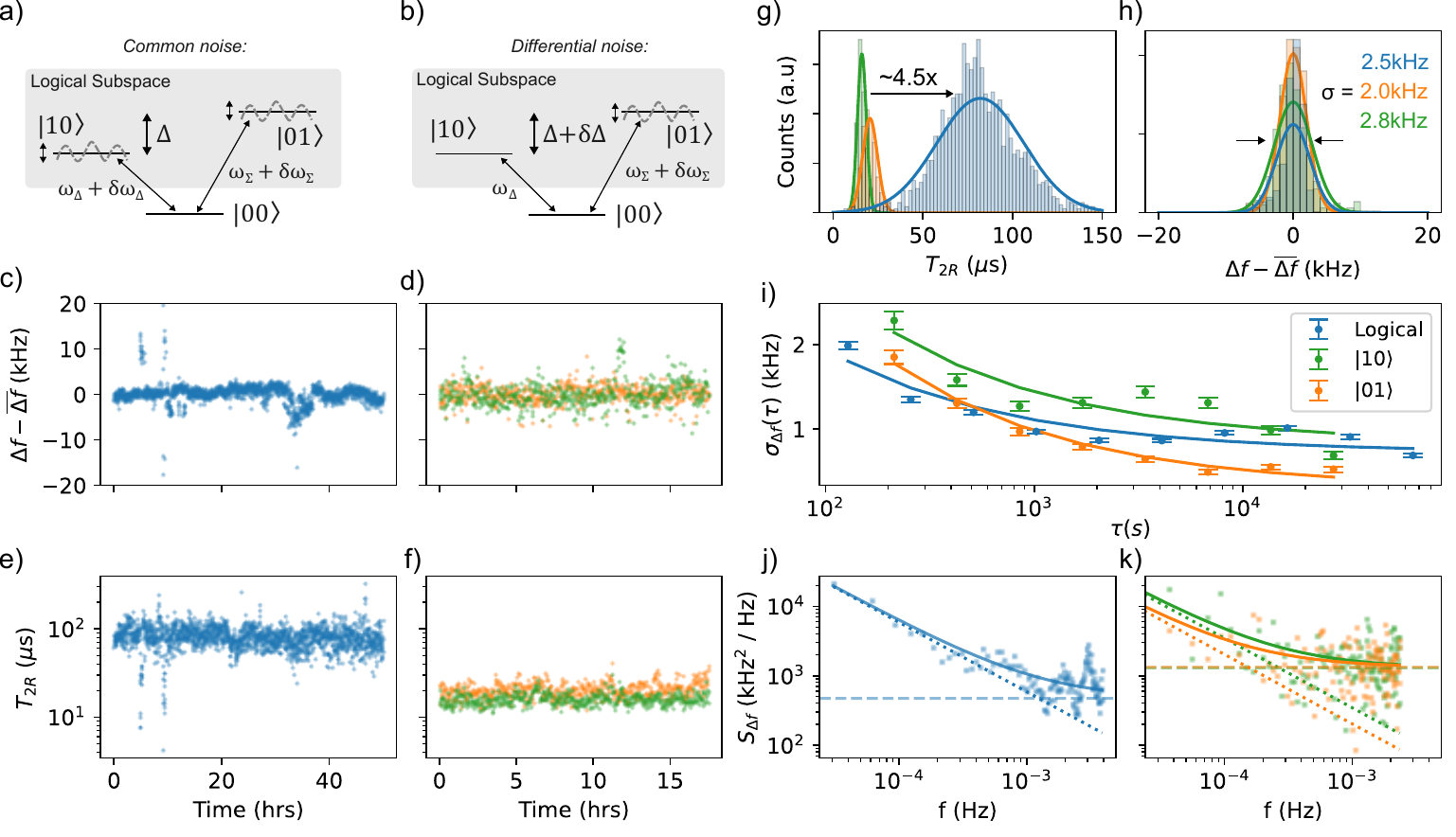}
\caption{\label{fig:ramsey_analysis}Ramsey interferometry measurements of logical encoded qubit, and physical modes of Q2. (a-b) Energy level diagrams of the dimon qubit with transition frequency fluctuations caused by (a) common noise, and (b) differential noise. (c-d) Extracted oscillation frequency deviation from mean frequency ($\Delta f - \overline{\Delta f}$) of the (c) logical qubit, and (d) Q-mode (orange) and D-mode (green) Ramsey measurement. Physical modes measurement interleaved over course of 17 hrs, logical qubit measured separately over 50 hrs. (e-f) Extracted Ramsey decay constant ($T_{2R}$) of the (e) logical encoded qubit and (f) physical modes. (g-h) Histograms of extracted parameters ($T_{2R}$, $\Delta f - \overline{\Delta f}$) of the logical qubit and physical modes. (i) Overlapping Allan deviation of frequency fluctuations. (j-k) Welch-method spectral density of frequency fluctuations of the logical Ramsey measurement (j), and physical modes (k). Data (crosses) is fit (solid line) to a noise model outlined in main text. We show extracted white noise amplitude (dashed line), in addition to extracted 1/f noise amplitude (dotted line).}
\end{figure*}

For any such noise sources that cause a fluctuation in both $\omega_Q$ and $\omega_D$, as shown in Fig. \ref{fig:ramsey_analysis} (a), the DDQ has a reduced sensitivity (or is entirely insensitive) to in terms of coherence. We describe these as \textit{common} noise sources, and physical examples include charge noise \cite{wills_spatial_2022}, quasiparticle tunneling \cite{de_graaf_two-level_2020}, and far-detuned two-levels systems (TLS) \cite{muller_towards_2019}. In each of these cases, the noise source fluctuates both dimon mode frequencies simultaneously, such that $\delta \Delta$ is small, or zero in perfectly symmetrical cases. As such, if common noise sources were a dominating cause of decoherence in this device, we would expect to observe an increase in coherence of the DDQ in the logical subspace compared to the physical modes.

Where the coherence of the DDQ does possess a sensitivity is to what we describe as \textit{differential} noise sources. These are noise sources that affect only one of the transition frequencies of the dimon, as shown in Fig. \ref{fig:ramsey_analysis} (b), such that $\delta \omega_Q \neq \delta \omega_D$. Examples of physical sources that can cause this kind of differential noise are frequency dependent control electronics noise, and nearly-resonant TLS \cite{ku_decoherence_2005, thorbeck_two-level-system_2023, carroll_dynamics_2022, schlor_correlating_2019}.

In the case of the DDQ, there are noise sources where the logical encoding gives a degree of protection of coherence, but the asymmetry of the device means there is still some sensitivity. Two prominent examples of this are Josephson energy fluctuations due to critical current or dielectric noise, and resonator photon shot noise dephasing \cite{zhang_suppression_2017}. For dephasing due to photon fluctuations in the resonator, we approximate the ratio of logical dephasing ($\Gamma_\phi^L$) to average physical mode dephasing ($\Gamma_\phi^{phys}$) as, 
\begin{equation}\label{eq:res_dephasing_ratio}
    \frac{\Gamma_\phi^{L}}{\Gamma_\phi^{phys}} = \frac{\delta\chi^2 \left( \kappa^2 + 4 \overline{\chi}^2 \right)}{\overline{\chi}^2 \left( \kappa^2 + 4\delta\chi^2 \right)},
\end{equation}
where $\overline{\chi} = (\chi_{QR} + \chi_{DR})/2$, and $\delta \chi = \chi_{QR} - \chi_{DR}$. With the measured parameters, we estimate the logical encoding to result in a reduction in dephasing rate of $42\%$, $11\%$ and $18\%$, for Q1, Q2 and Q3 respectively.

The Josephson junctions are also a source of asymmetry in the fixed frequency DDQ architecture, since nanofabrication methods result in a statistical spread of junction parameters. By considering fluctuations in the detuning between the physical modes of the dimon, we calculate the dephasing rate $\Gamma_\phi^L \propto \partial\Delta / \partial E_{J1,2} \approx (1-\sqrt{r})/\Delta$, where $r = E_{J1}/E_{J2}$ is the ratio of the two junction energies. In the case of perfectly symmetrical junctions ($E_{J1} = E_{J2}$), the DDQ is protected from fluctuations in junction energies, since both junctions participate equally in each mode of the dimon. From room temperature resistance measurements, we infer a junction ratio of $r = 0.963, 0.959 $ and $0.931$, for Q1, Q2 and Q3 respectively.

In both of these physical noise sources, the device can be further engineered with symmetry protection, through methods of symmetric qubit-resonator coupling and more favorable $\delta\chi/\overline{\chi}$ ratios for resonator photon shot noise dephasing protection, or Josephson junction fabrication postprocessing techniques \cite{hertzberg_laser-annealing_2021, pappas_alternating_2024, balaji_electron-beam_2024, kennedy_tuning_2025} for critical current noise protection. We also postulate that increasing the coupling strength between the modes, and hence the detuning, $\Delta$, reduces sensitivity to Josephson junction asymmetry, where post-fabrication tuning methods are not viable.

To further probe the dynamics of the Ramsey coherence measurements, we perform more detailed statistical analysis and methods of decoherence benchmarking \cite{burnett_decoherence_2019} on the time series data of Q2. In Fig. \ref{fig:ramsey_analysis} (c-f) we show the dynamics of the extracted frequency deviation from the mean oscillation frequency, $\overline{\Delta f}$, (c-d), and decay constants, $T_{2R}$, (e-f) over the course of an extended set of repeated measurements. The logical qubit is measured independently over a separate time period, and the physical mode measurements are interleaved with each other. In both the D-mode and logical qubit time series traces, we observe sudden jumps in frequency, which in the case of the logical qubit, align with drops in coherence. By contrast, we do not observe such jumps in the Q-mode traces, and the D-mode jumps do not correlate with drops in physical mode coherence. In previous studies, these jumps are attributed to TLS switching behavior, and often identified by observing similar telegraph-like noise.

In Fig. \ref{fig:ramsey_analysis} (g-h), we show histograms of the extracted parameters $T_{2R}$ and $\Delta f - \overline{\Delta f}$, for both the physical modes and the logical encoded qubit, from Fig. \ref{fig:ramsey_analysis} (c-f). In the case of the exponential decay constant $T_{2R}$, we see an improvement in the logical qubit coherence by factor of approximately 4.5x, larger than the 1-2x improvement predicted by Eqn. \ref{eq:res_dephasing_ratio}. Whilst the error-detected coherence has improved, the frequency stability of the device has not significantly changed, as shown by the widths of the histograms in Fig. \ref{fig:ramsey_analysis} (h). The causes of this, and the impact it has on the architecture moving forward require further investigation.

An overlapping Allan deviation analysis of the frequency fluctuations gives an insight into the stability of the measured values, as well as the noise processes impacting them. We show this analysis in Fig. \ref{fig:ramsey_analysis} (i), plotting both the physical and logical extracted frequency deviations, and fitting them to a model of $\sigma_{\Delta f}(\tau) = (B/2)^{1/2} \tau^{-\frac{1}{2}} + (2\ln(2)A)^{1/2}$, where $B$ describes the amplitude of white noise processes, and $A$ describes the amplitude of $1/f$ noise processes \cite{burnett_decoherence_2019}. Where the Q-mode measurement follows the expected model, both the D-mode and logical frequency fluctuations show peaks at long time scales ($\tau = 5e3 - 5e4$ seconds), which are not described by the noise model. As with previous decoherence benchmarking investigations \cite{burnett_decoherence_2019}, these peaks are commensurate with slow Lorentzian noise processes attributed to TLS switching behaviors, and can be observed in the time series traces of Fig. \ref{fig:ramsey_analysis} (c) and (d) as large infrequent jumps. The presence of these peaks in both the logical qubit and one of the physical modes suggests a differential noise source to which the logical encoding possesses sensitivities. 

Finally, in Fig. \ref{fig:ramsey_analysis} (j-k), we show a Welch-method spectral density analysis of the frequency deviations, extracted from the Ramsey measurement of both the logical encoded qubit (j), and physical modes (k). We fit this is a simplified noise model of $A/f + B$, finding $1/f$ noise amplitudes of $A = 5.9\times10^5$ Hz$^2$ for the logical qubit, and $A = 3.5\times10^5$ Hz$^2$ and $A = 2.0\times10^5$ Hz$^2$ for the D and Q modes respectively. Whilst the $1/f$ noise amplitude of the logical qubit is higher than the physical modes, the white noise parameter, $B$ is lower at $B = 0.5 \times 10^6$ Hz$^2$/Hz, compared to $B = 1.3 \times 10^6$ Hz$^2$/Hz for both physical modes.

As shown in Section \ref{sec:coherence_stats}, we observe a significant improvement in both bit-flip and phase-flip rates in the error-detected subspace, whereas the Ramsey coherence does not show as dramatic an improvement. In addition, where error-detected $T_1^L$ and $T_{2E}^L$ measurements show second order decay profiles, the Ramsey measurement follows the form of an exponentially decaying oscillation, indicative of Lorentzian noise processes. From our analysis we postulate that quasistatic differential noise source is the leading source of error in this implementation of logical encoding and error-detection. Furthermore, the improvement in logical coherence suggests a common noise source is dominant in the physical qubit modes, such as a far detuned TLS, or the physical mechanisms described previously.

The precise source of this noise source affecting the logical encoding requires further investigation. Whilst our analysis suggests protection from physical noise sources such as critical current noise and resonator photon shot noise, we do not observe a correlation between qubits with the more favorable $\delta \chi / \overline{\chi}$ or junction symmetry ratios, $r$, and logical qubit coherence. One potential candidate is a nearly resonant TLS that has a low-frequency switching-like behavior, causing the infrequent jumps in both frequency detuning $\Delta f$ and $T_{2R}^L$, over the timescale of hours as indicated in the overlapping Allan deviation analysis in Fig. \ref{fig:ramsey_analysis} (i). This is consistent with previous observations of telegraph noise in dual-rail encoded superconducting qubit architectures with flux tunability \cite{levine_demonstrating_2024}, as well as decoherence benchmarking measurements in fixed frequency transmon qubits \cite{burnett_decoherence_2019}. Advanced nanofabrication and processing methods can show a reduction in the density of these two-level systems and fluctuators \cite{pappas_alternating_2024}, and we propose the DDQ as a tool to further probe the efficacy of these methods. We suggest using the error-detected coherence metrology alongside more detailed methods such as noise spectroscopy analysis \cite{gupta_expedited_2025, wise_using_2021} would allow for a more complete picture of the fundamentals of noise and decoherence in superconducting qubits.

\section{\label{sec:conclusion}Conclusion}
In this work we have demonstrated the encoding of an error-detected logical qubit in a fixed-frequency multimode transmon qubit using dual-rail encoding. We have demonstrated protocols for reading out the logical state; performing an end-of-line amplitude damping error-detection and, through the process of error-detection and postselection, we have shown how bit-flip and phase flip error rates in the logical subspace are greatly reduced in comparison to the constituent physical modes of the device. Finally, we propose this device has utility to investigate higher order effects of noise sources on coherence of superconducting quantum devices, and as a tool for evaluating noise mitigation strategies. Understanding these noise sources can allow future generation qubits to overcome current limitations and operate with significantly reduced error rates.

\begin{acknowledgments}
We extend our thanks to the entire OQC Team for their contributions to the quantum computing stack, which was instrumental in this work. The authors thank Jonathan Burnett, Oscar Kennedy, Ailsa Keyser, Boris Shteynas, and Todd Tilma for reviewing this manuscript.
\end{acknowledgments}

\appendix

\section{\label{sec:device_params}Dimon Device Parameters}
In Table \ref{table:1} we show the measured parameters of the DDQ device. Transition frequencies, anharmonicities, and cross-Kerr shifts are obtained via spectroscopic measurements. Energy relaxation time ($T_1$), Hahn-echo ($T_{2E}$) and Ramsey dephasing ($T_{2R}$) times are obtained via conventional time domain measurements as outlined in \cite{krantz_quantum_2019}. Coherence metric values reported are medians.

\begin{table*}
\caption{Device Parameters}
\label{table:1}

\begin{tabularx}{0.75\textwidth}{p{6.5cm} p{2.5cm} p{2.5cm} p{2.5cm} } 
 \hline
 Parameter & Q1 & Q2 & Q3\\ [0.5ex] 
\hline
 \textbf{LC Resonator} &   \\ 
 Frequency $\omega_R/2\pi$ [GHz] & 9.997 & 10.436 & 10.432\\
 Linewidth $\kappa_R/2\pi$ [MHz]& 0.81 & 1.01 & 0.79  \\
 Dispersive Shift $2\chi_{DR}/2\pi$ [MHz]& 1.42 & 1.63 & 1.33   \\
 Dispersive Shift $2\chi_{QR}/2\pi$ [MHz]& 1.86 & 2.18 & 2.06   \\
  &   \\ 
 \textbf{D Mode}  &   \\ 
 Transition Frequency  $\omega_D/2\pi$ [GHz]&   4.707 & 4.353 & 4.235\\
 Anharmonicity  $\alpha_D/2\pi$ [MHz]&   -136 & -126 & -120\\ 
 $T_1$ [$\mu$s]&   63.6 & 81.2 & 70.9\\ 
 $T_{2E}$ [$\mu$s]&   63.4 & 74.3 & 41.1 \\ 
 $T_{2R}$ [$\mu$s]  &  15.6 & 15.9 & 25.2 \\ 
   &   \\ 
 \textbf{Q Mode}  &   \\ 
 Transition Frequency  $\omega_Q/2\pi$ [GHz]&   5.468 & 5.403 & 5.276\\
 Anharmonicity  $\alpha_Q/2\pi$ [MHz]&   -156 & -168 & -164\\ 
 $T_1$ [$\mu$s]&   65.3 & 55.4 & 55.8\\ 
 $T_{2E}$ [$\mu$s]&   87.1 & 57.2 & 78.4\\ 
 $T_{2R}$ [$\mu$s]&   16.6 & 20.3 & 18.4\\ 
    &   \\ 
 Cross-Kerr Shift $\eta/2\pi$ [MHz]&   -283 & -270 & -270\\ 
    &   \\ 
 \textbf{Logical Qubit}  &   \\ 
  Mode Detuning  $\Delta/2\pi$ [GHz]&   0.761 & 1.050 & 1.041\\
  1/Bit-flip rate $T_1^L$ [ms]&   4.47 & 3.86 & 1.03\\
  1/Phase-flip rate $T_{2E}^L$ [ms]&   0.85 & 0.68 & 0.75\\
  Ramsey Decay $T_{2R}^L$ [ms]&   0.05 & 0.08 & N/A\\

 \hline
\end{tabularx}
\end{table*}

\section{\label{sec:time_series_data}Time Series Data}
\begin{figure*}
\includegraphics[width=0.95\textwidth]{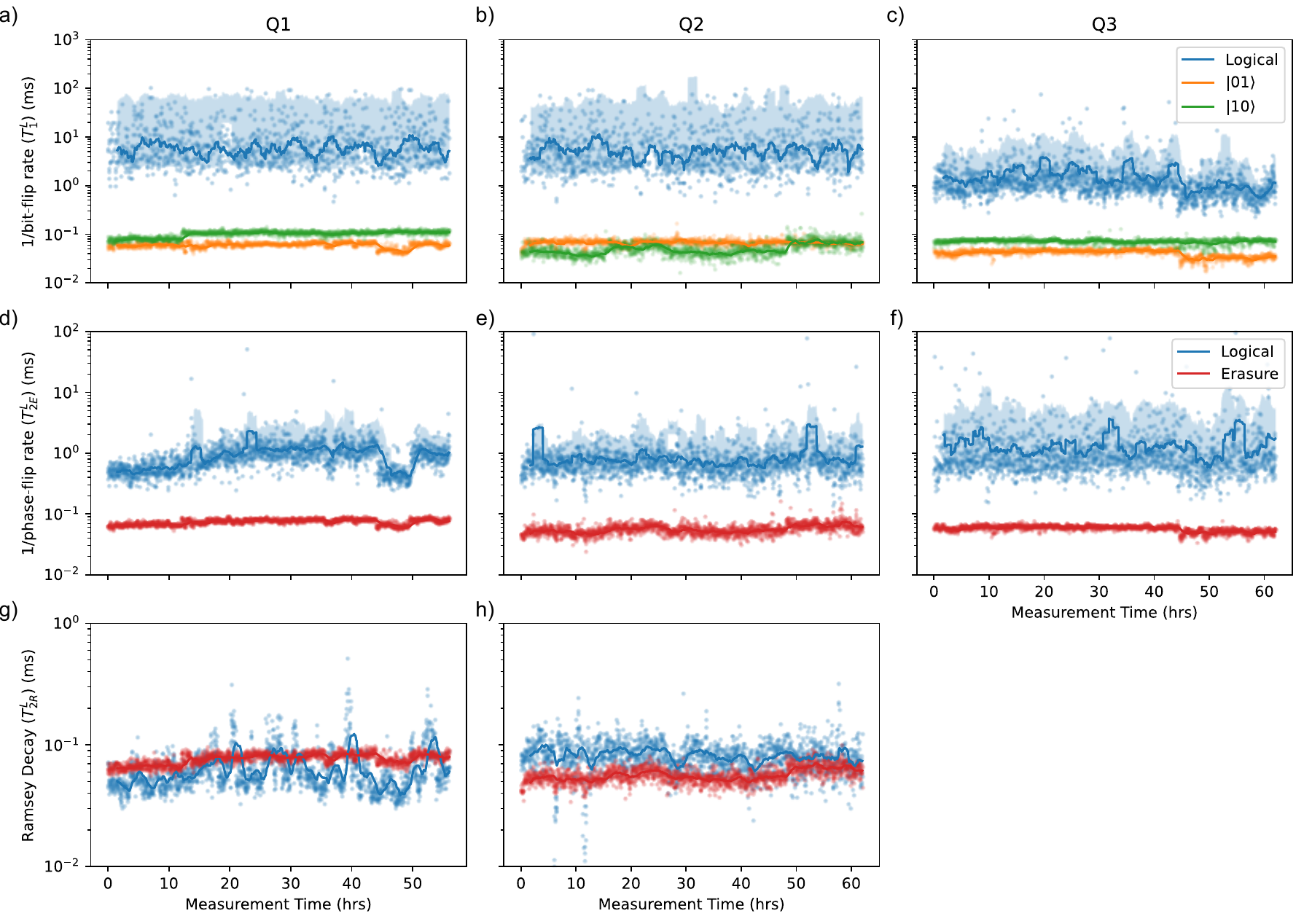}
\caption{\label{fig:coh_time_traces}Time series traces of error-detected coherence metrics. (a-c) Extracted logical qubit $T_1^L$ (blue) of Q1, Q2 and Q3, measured over the course of over 50 hours. Semi-transparent points show extracted values from individual traces, with solid line showing a moving average with a window size of 50 traces, corresponding to approximately 1.5 hours. Shaded blue area shows bounds generated by statistical bootstrapping methodology, outlined in main text. Physical $T_1$ of the $|01 \rangle$ ($|10 \rangle$) state shown in orange (green), with individual extracted values shown by semi-transparent points, and moving average indicated by the solid line. (d-f) Extracted logical coherence time $T_{2E}^L$ (blue) of Q1, Q2 and Q3. Individual extracted values, bounds generated by statistical bootstrapping, and moving averages shown as in logical $T_1^L$ plots. Extracted erasure rate shown in red. (g-h) Extracted error-detected Ramsey decay constant $T_{2R}^L$ (blue) of Q1, Q2, and erasure rate (red). Individual extracted values shown as semi-transparent points, and moving averages indicated by solid lines.}
\end{figure*}

We repeatedly measure the logical coherence metrics outlined in the main text over a period of over 50 hours to build statistics. Q2 and Q3 are measured simultaneously, whereas Q1 is measured over a separate 50 hour period. Measurements of bit-flip rate, phase-flip rate, and Ramsey coherence are interleaved, with a total of 1750 data points for each metric measured across the three devices. The fitted values as a function of time are shown in Fig. \ref{fig:coh_time_traces}.

Upper and lower bounds of the fits are obtained by method of statistical bootstrapping. For each trace that is measured a fit value is obtained as well as residual values, giving the deviation of each point in the trace from the ideal value. We take the ideal trace for each fit value, and add samples of the previously extracted residual values to generate 250 simulated noisy traces. These simulated traces are then fit, and the bounds plotted in Fig. \ref{fig:coh_time_traces} are obtained from the upper and lower 5\% quantiles from the resulting distributions of extracted fit values.

\bibliography{bibliography}

\end{document}